\documentclass[aps,prb,reprint]{revtex4-2}

\usepackage{latexsym,graphicx,graphics}
\usepackage{appendix}
\usepackage{amsmath,epsfig}

\newcommand{\alphab}{\bar{\alpha}}
\newcommand{\betab}{\bar{\beta}}
\newcommand{\kob}{\bar{k}_0}

\newcommand{\rone}[1]{\boldsymbol{#1}}
\newcommand{\rtwo}[1]{\underline{\underline{\boldsymbol{#1}}}}

\begin{document}

\title{Nonreciprocal surface plasmons in angularly varying, magnetized, metasurface tubes}

\author{Yarden Mazor}
 \affiliation{School Of Electrical Engineering, Tel Aviv University, Tel Aviv, 69978, Israel}
 \email{Yardenm2@tauex.tau.ac.il}




\begin{abstract}
We analytically and numerically study the nonreciprocal surface waves guided by a magnetized metasurface tube. When applying the magnetic bias perpendicularly to the cylinder axis, the conductivity profile cross-section is nonuniform, which enables us to obtain pronounced nonreciprocal modes with different field distributions for propagation in opposite directions due to the coupling between different values of orbital angular momentum. We show that this property can be leveraged for directional power delivery when changing the source location, and we study the isolation ratio as a function of different parameters. The simple, continuous dependence of the isolation on the direction of magnetization allows simple yet robust control over the wave propagation properties.
\end{abstract}
\maketitle


\section{Introduction}
Cylindrical structures are among the most common waveguide geometries being used. Prominent examples include coaxial waveguides and hollow metallic tubes \cite{collin1992foundations}. The dispersion of guided waves can be controlled by varying the cylinder size (potentially also the internal cylinder in a coaxial geometry) and the distribution and properties of the filling material. These waveguides have many uses, such as communications \cite{collin1992foundations}, open-ended aperture antennas \cite{kraus_antennas_2002,Papas}, material sensing probes, and more. 

More recently, incorporating metamaterials and metasurfaces within a cylindrical geometry was examined as a platform for various applications. Scattering from cylindrical metasurface structures was studied in \cite{YakovlevCloaking} to achieve cloaking, and in \cite{grbic2021analysis}, the authors show how to systematically analyze and synthesize a layered cylindrical structure to achieve different scattering functionalities. Another venue is tailoring guided wave properties in cylindrical waveguides using metasurfaces coating the waveguide walls or serving as the waveguide themselves. This is a particularly interesting venue since cylindrical guided modes exhibit complex polarization states, helicity, and spin-OAM coupling \cite{BliokhHelicity}. This was pioneered in \cite{PierceHelix}, where a helically slotted PEC cylinder was studied. A modal theory for such structures with general electric impedance walls was formulated in \cite{RaveuCylindricalTheory}. In \cite{IyerCylindricalMetamaterial}, a metamaterial lining was used to control the cut-off and modal properties, and in \cite{mazor2019angular} it was shown how general bianisotropic walls can be used to tailor the modal dispersion, field distribution, and spin. A step forward was taken in \cite{IyerBarker2023}, where the authors discuss an actual implementation and consider realistic constraints and the arising dispersion effects. Such structures also appear in more natural systems, where the most prominent example is carbon nanotubes. The electrodynamics of cylindrical graphene-based structures was first studied in \cite{SlepyanCNT}, where the appropriate boundary conditions were formulated, and the dispersion of guided surface waves was carefully mapped. In the last decade, several groups have fabricated and measured artificial cylindrical graphene structures, for enhancing light-matter interaction \cite{GrapheneLabOnRod2014,GraphenePHCfiber2019}, four-wave mixing \cite{GrapheneFWM2015} and all-optical modulation \cite{GrapheneModulator2015}. Multilayered folding is shown in \cite{GrapheneRoll2014}, and additional applications are reviewed in \cite{Silica2DReview2021}.

An equally important ingredient of this work is the nonreciprocal propagation of waves. Nonreciprocal waveguides have many applications, such as routing and sorting of signals, and serving as building blocks in the design of nonreciprocal components such as isolators and circulators \cite{EMnonreciprocity,MicrowaveNonreciprocity}. To obtain significant nonreciprocal behavior, such as different field distributions for oppositely propgating fields which can be leveraged for isolation \cite{collin1992foundations} (e.g., asymetrically ferrite loaded waveguides), or one-way guiding, one has to incorporate nonreciprocity with some form of structural asymmetry \cite{Fan2008nonrec,GroupSym}. In the context of tube-like structures, \cite{alu2015electrically} studied guided waves in several cylindrical geometries incorporating graphene, where electrical bias (controlling the chemical potential $\mu_c$) and magnetic bias were used as control parameters to manipulate the wave characteristics. Importantly, this work also incorporated nonreciprocity since the magnetic field bias gives rise to an asymmetric conductivity tensor of the graphene layer. Nevertheless, this nonreciprocity did not visibly manifest in the modal dispersion or field distribution due to the highly "symmetric" magnetization scheme considered - a radial magnetic field bias. 

In this work, we would like to examine the nonreciprocal wave propagation on a magnetized cylindrical impedance surface, such as graphene. Unlike other examples, where the structural asymmetry requires ingomogeneous cross section materials or scatterer parameters, here it stems from the magnetization scheme itself. By applying a simple magnetic bias perpendicular to the waveguiding surface (represented by a thin impedance sheet), an azimuthally dependant conductivity tensor arises. Therefore, the magnetization itself provides both the nonreciprocity and the spatial inhomogeneity. This, in turn, creates coupling between different values of orbital angular momentum (OAM) of the guided wave components. The nonreciprocity makes this coupling asymmetric, giving rise to complex nonreciprocal waves composed of a superposition of different cylindrical eigenfunctions (each with its characteristic OAM), resulting in different field distributions modes propagating in opposite directions. This will be shown to enable directional excitation using simple dipolar sources as a function of the source location and spin-momentum coupling. 
\section{Formulation}
Let us start from the basic formulation of the problem of waveguiding by thin cylindrical shells. Since the waves propagate essentially in free space (with the addition of the impedance/surface susceptibility boundary condition), we can represent them as a superposition of the "standard" TE and TM modes
\begin{subequations}
\begin{equation}
    E_z=
    \begin{Bmatrix}
    A_n^iI_n(\alpha r) & ;r<a \\
    A_n^oK_n(\alpha r)  & ;r>a
    \end{Bmatrix}
    e^{-j\beta z}e^{-jn\varphi},
    \label{eq:Ezdef}
\end{equation}
\begin{equation}
    H_z=
    \begin{Bmatrix}
    B_n^iI_n(\alpha r) & ;r<a \\
    B_n^oK_n(\alpha r)  & ;r>a
    \end{Bmatrix}
    e^{-j\beta z}e^{-jn\varphi},
    \label{eq:Hzdef}
\end{equation}
\end{subequations}
with $\beta^2=\alpha^2+k_0^2$, and $k_0=\omega\sqrt{\mu\epsilon}$ and $I_n,K_n$ are the modified Bessel functions of the first and second kind. The other field components are given can be directly derived from equations (\ref{eq:Ezdef}),({\ref{eq:Hzdef}}), and are given in appendix \ref{VaccumFields} for completeness.

In the present problem, the waves are guided by a tube of radius $R$, which satisfies the surface impedance boundary condition
\begin{equation}
 \hat{\boldsymbol{r}}\times\left(\boldsymbol{H}_{out}-\boldsymbol{H}_{in}\right)_{r=R}=\boldsymbol{J}_s=\rtwo{\sigma}\boldsymbol{E}_{tan,r=R},
    \label{eq:Impedance_BC}
\end{equation}
where $\rtwo{\sigma}$ is a $2\times 2$ matrix, that can be written explicitly as 
\begin{equation}
\rtwo{\sigma}=\begin{bmatrix}
\sigma_{\varphi\varphi} & \sigma_{\varphi z} \\
\sigma_{z\varphi} & \sigma_{z z},
\end{bmatrix}
\end{equation}
and $\rone{E}_{tan}$ is the electric field components tangent to the tube. In addition to this, since the cylindrical surface only hosts electric surface currents, the tangential electric fields need to be continuous
\begin{equation}
    E_z(R^+)=E_z(R^-)\;,\;E_{\varphi}( R^+)=E_{\varphi}(R^-),
    \label{eq:EContinuity}
\end{equation}
where $()^+ and ()^-$ indicate sampling of the fields on the surface of the tube, on the outer and inner side, respectively. Using this condition, the field representation on the surface of the tube can be reduced to rely on two coefficients $A_n^i,\bar{B}_n^i=\eta_0B_n^i$, (see appendix \ref{VaccumFields}). If we now define
\begin{equation}
    \rone{D}_n=\begin{pmatrix}
    A_n^i \\ \bar{B}_n^i
    \end{pmatrix},
    \label{Ddefinition}
\end{equation}
we can represent the operations required for substitution into equation (\ref{eq:Impedance_BC}) as
\begin{subequations}
\begin{equation}
    \rone{E}^o_{n,tan}(r=R)=\rone{E}^i_{n,tan}(r=R)=\rtwo{M}_{E,n}\rone{D}_n,
    \label{Eboundary_ME}
\end{equation}
\begin{equation}
    \hat{\boldsymbol{r}}\times(\rone{H}^o_n-\rone{H}^i_n)_{(r=R)}=\rtwo{M}_{H,n}\rone{D}_n, 
    \label{Hboundary_MH}
\end{equation}
\end{subequations}
with
\begin{equation}
    \rtwo{M}_{E,n}=\begin{bmatrix}
    \frac{n\betab}{\alphab^2}I_n & -\frac{j\kob}{\alphab}I_n' \\
    I_n & 0 
    \end{bmatrix}
    \;,\rtwo{M}_{H,n}=\begin{bmatrix}
    0 & -\frac{1}{\alphab K_n'} \\
    -\frac{j\kob}{\alphab K_n} & \frac{n\betab}{\alphab^3 K_n'}
    \end{bmatrix}.
    \label{MEMHdefinition}
\end{equation}
To shorten the notation, we will omit the argument of the modified Bessel function $I_n,K_n$ and their derivatives $I_n',K_n'$ when we refer to sampling these on the surface of the tube. In addition, we have defined $\alphab=\alpha R\;,\;\betab=\beta R\;,\;\kob=k_0 R$.
Now, if we substitute this representation into the boundary condition given in equation (\ref{eq:Impedance_BC}) and assume the conductivity $\rtwo{\sigma}$ is independant of $\varphi$, we obtain the eigenmode equation
\begin{equation}
    \left(\eta_0\rtwo{\sigma}\,\rtwo{M}_{E,n}-\rtwo{M}_{H,n}\right)\rone{D}_n=0.
    \label{EigenmodeSingleN}
\end{equation}
For this equation to have a non-trivial solution for the coefficient vector $\rone{D}_n$, the determinant of the bracketed matrix must vanish, which yields the dispersion relation as a determinant of a $2\times 2$ matrix. For any value of $n$ (the OAM of the mode) and angular frequency $\omega$ we substitute, we will obtain $\beta(\omega)$ - the dispersion relation of the $n$'th mode. Substituting this value back will yield the modal coefficients $A_n^i,\bar{B}_n^i$, forming a mixed TE/TM mode, with the amplitude of each constituent depending on the conductivity. 

Based on this, our goal now would be to describe the waves propagating in a system where $\rtwo{\sigma}$ is a function of the azimuthal angle $\varphi$, generally expressed as $\rtwo{\sigma}(\varphi)$. We start by describing all the relevant quantities $\rone{E},\rone{H},\rtwo{\sigma}$ using their Fourier series in $\varphi$. For each specific mode $n$ we use the matrices defined in equations (\ref{Eboundary_ME}),(\ref{Hboundary_MH}),(\ref{MEMHdefinition}) to represent the fields using the corresponding coefficients $\rone{D}_n$, yieding
\begin{subequations}
\begin{equation}
    \begin{pmatrix}
    E_{\varphi}^i \\ E_z^i
    \end{pmatrix}
    =\sum_{n=-\infty}^{\infty}\rtwo{M}_{E,n}\rone{D}_ne^{-jn\varphi},
\end{equation}
\begin{equation}
    \hat{\boldsymbol{r}}\times
    \begin{pmatrix}
    H_{\varphi}^o-H_{\varphi}^i \\ H_z^o-H_z^i
    \end{pmatrix}
    =\sum_{n=-\infty}^{\infty}\rtwo{M}_{H,n}\rone{D}_ne^{-jn\varphi},
\end{equation}
\end{subequations}
and for the condictivity
\begin{equation}
    \eta\rtwo{\sigma}=\sum_{m=-\infty}^{\infty}\rtwo{Y}_me^{-jm\varphi}.
\end{equation}
When $\rtwo{\boldsymbol{\sigma}}=\rtwo{\boldsymbol{\sigma}}(\varphi)$, there is coupling between different components with different $n$, through the additional OAM provided by the geometrical structure of $\rtwo{\sigma}$, and we can derive the modal equation by substituting these expansions into equation (\ref{eq:Impedance_BC}), 
\begin{equation}
    \sum_{n=-\infty}^{\infty}\rtwo{Y}_{\ell-n}\rtwo{M}_{E,n}\rone{D}_n-\rtwo{M}_{H,\ell}\rone{D}_{\ell}=0\;\;,\;\;\forall \ell,
    \label{EigenmodeEq}
\end{equation}
where each value of $\ell$ represents a row in a $2\times 2$ infinite block matrix. This can be reformulated into a matrix equation, as shown in appendix \ref{App:MatrixInf}.

For this system to have a nontrivial solution, the determinant of the matrix (\ref{masterMatrix}) must vanish, giving the dispersion equation for the propagating modes. Naturally, since different values of $n$ are coupled through the angle-dependent conductivity of the tube, the modes will be composed of many azimuthal harmonics excited and propagating together. This is in addition to the natural TE/TM coupling in higher order modes propagating on impedance cylinders, giving rise to hybrid-type modes that can yield interesting patterns of spin and helicity \cite{mazor2019angular}.

For the sake of simplicity, We will limit our discussion here to conductivity matrices of the form
\begin{equation}
    \rtwo{\sigma}(\varphi)=\rtwo{\sigma}_0+\rtwo{\sigma}_1\cos\varphi=\rtwo{\sigma}_0+\frac{1}{2}\left(\rtwo{\sigma}_1e^{j\varphi}+\rtwo{\sigma}_1e^{-j\varphi}\right).
    \label{SigmaPhi}
\end{equation}
This would allow us to simplify equation (\ref{EigenmodeEq}) and its matrix representation (\ref{masterMatrix}) to a tridiagonal block form.
In this case, one could formally express the dispersion relation in a matrix continued fraction form, in a process similar to the one described in \cite{OlinerSinusoidalReactance} 
\begin{widetext}
\begin{multline}
    \rtwo{Y}_1^{-1}\left[-\rtwo{Y}_0+\rtwo{G}_n+\left[-\rtwo{Y}_0+\rtwo{G}_{n-1}+\left[ \rtwo{Y}_0+\rtwo{G}_{n-2}+[...]^{-1} \right] ^{-1}\right]^{-1} \right]=\\
    \left[ \rtwo{Y}_1^{-1}\rtwo{Y}_0-\rtwo{Y}_1^{-1}\rtwo{G}_{n+1}+\left[ \rtwo{Y}_1^{-1}\rtwo{Y}_0-\rtwo{Y}_1^{-1}\rtwo{G}_{n+2}+[...]^{-1}\right]^{-1}\right]^{-1}
\end{multline}
\end{widetext}
With $\rtwo{G}_n=\rtwo{M}_{H,n}\rtwo{M}_{E,n}^{-1}$, and we have assumed $\rtwo{Y}_1=\rtwo{Y}_{-1}$ as in our simplified case.

Since we would like to study nonreciprocal propagation, it is worth noticing that when the guiding structure is reciprocal (satisfying  $\rtwo{\sigma}=\rtwo{\sigma}^T$), inverting the propagation direction would result in time-reversed modal fields. For a specific mode, the modal fields will be transformed as $\rone{E}(-\beta)=\rone{E}(\beta)^*$,$\rone{H}(-\beta)=-\rone{H}^*(\beta)$. However, this is no longer the case when the conductivity tensor does not satisfy reciprocity. In many cases, merely violating the reciprocity of the guiding structure is not enough for visible effects of non-reciprocity to manifest in the guided modes (such as a non-symmetric dispersion curve and different field profiles for oppositely propagating waves). This is one of the reasons that \cite{alu2015electrically} did not observe any significant nonreciprocal propagation and excitation effects, although studying a nonreciprocal structure. However, we will see that more significant nonreciprocal effects are possible in our case due to the explicit geometrical asymmetry arising from the $\varphi$ dependence of the conductivity matrix.
\subsection{Transversely magnetized gyrotropic tube}
\begin{figure}[t]
\begin{center}
\noindent
  \includegraphics[width=.85\columnwidth]{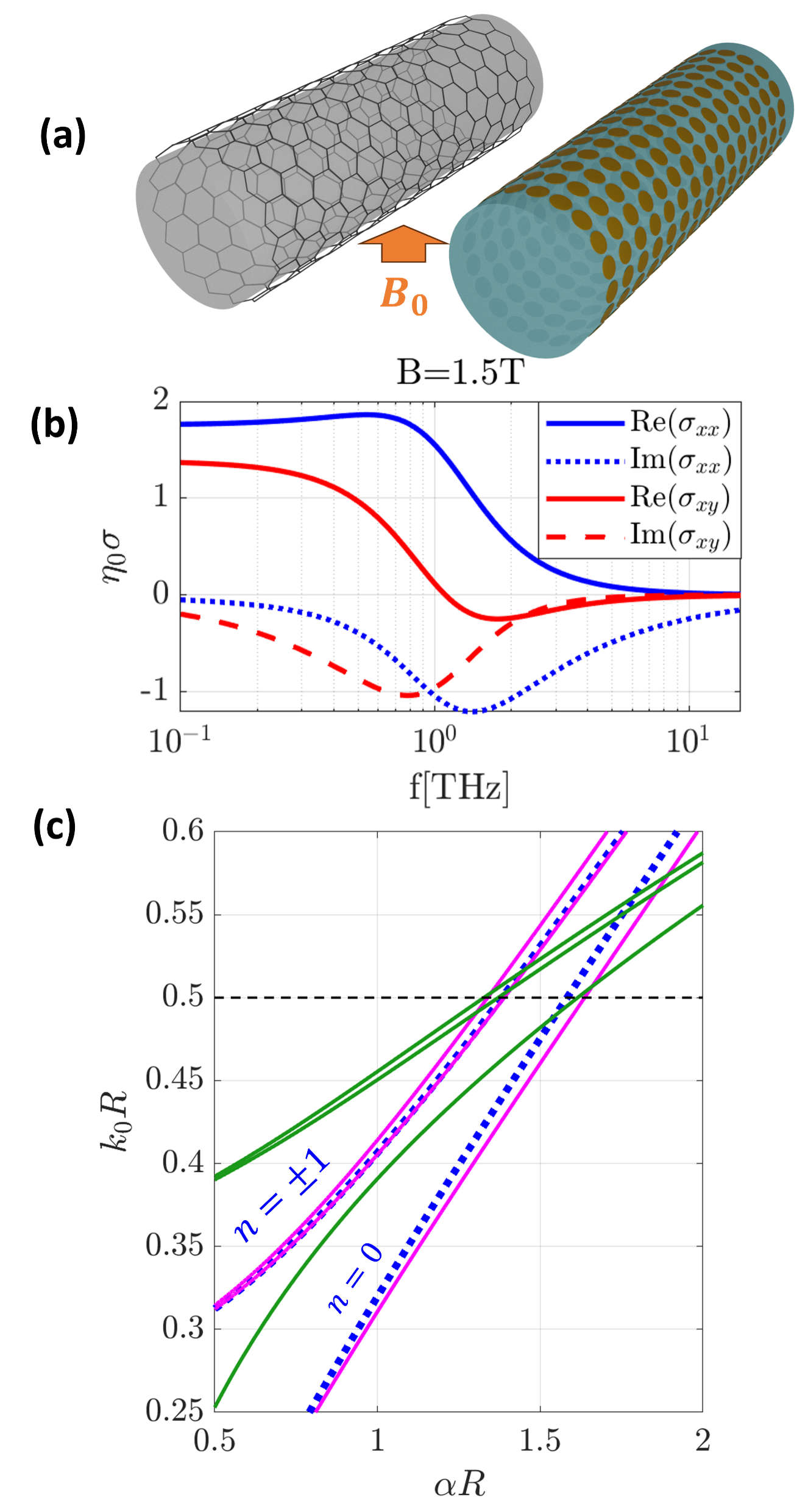}
  \caption{(a) Possible geometries for magnetizable cylindrical surface waveguides. (b) Graphene conductivity under a magnetization of $B_0=1.5T$. (c) Surface wave dispersion. Blue - no magnetization, red - with magnetization, green - accounting for the dispersion of graphene conductivity.}
  \label{fig1}
\end{center}
\end{figure}

The case we would like to focus on is where the conductivity of the cylinder can be expressed as
\begin{equation}
\rtwo{\boldsymbol{\sigma}}(\varphi)=\frac{1}{\eta_0}\begin{pmatrix}
jX & X_o\cos\varphi \\
-X_o\cos\varphi & jX
\end{pmatrix}.
\label{eq:sigmaMat}
\end{equation}
Although this choice seems quite specific, such a system naturally arises if we envision a Graphene tube, and magnetize it perpendicularly to the cylinder, as shown in figure \ref{fig1}(a) (in a setup similar to \cite{GrapheneLabOnRod2014,IEEEphotonicsGMF} for instance). Since the magnitude of the off-diagonal conductivity term is proportional to the magnetic field component perpendicular to the surface, and assuming the external magnetic field is uniform, say $\boldsymbol{B}=B_0\hat{x}$, we will obtain such a conductivity matrix \cite{sounas2012gyrotropy}. Due to the magnetization, the diagonal terms would also depend slightly on $\varphi$. However, for magnetic fields up to $~1.5T$ this dependence is relatively weak ($<10\%$ deviation from the value without magnetic field). Therefore, we will neglect this dependence, taking $X$ in equation (\ref{eq:sigmaMat}) independent of $\varphi$. This variation can be accounted for straightforwardly using the same formulation. Cylindrical arrangements of plasmonic patches or other particles are also expected to yield a similar conductivity structure (also shown in figure \ref{fig1}(a)). Extreme wave propagation effects using plasmonic particles and patches was demonstrated in \cite{HyperbolicPatches2015}, and nonreciprocity was shown in a planar setup in \cite{Metaweaves2014}. Here, we do not require the particle anisotropy (implemented through the usage of elongated ellipsoidal pathces) since the spatial symmetry breaking required for significant nonreciprocity occurs due to the cross-section inhomogeneity in our case.
To analyze this system using the developed formulation, we represent the conductivity tensor using its $\varphi$ Fourier transform. The matrix coefficients are
\begin{equation}
    \rtwo{Y}_0=\begin{pmatrix}
    jX & 0 \\
    0 & jX
    \end{pmatrix},\;\;
    \rtwo{Y}_{-1}=\rtwo{Y}_1=\frac{1}{2}\begin{pmatrix}
    0 & -X_o \\
    X_o & 0
    \end{pmatrix}.
\end{equation}
Once substituted into equation (\ref{EigenmodeEq}) (or \ref{masterMatrix}), we obtain a tridiagonal block system of equations. Although the magnetic bias makes this system nonreciprocal, we know from previous works that nonreciprocity can manifest in several ways, which are often quite subtle. 

The conductivity matrix elements for Graphene are shown in figure \ref{fig1}(b). The parameters we take are the Fermi velocity $v_F=10^6m/s$, the chemical potential $\mu_c=0.35V$, and a biasing magnetic field of $B_0=1.5T$. Since the response of Graphene can be quite lossy in the frequency regions where the gyrotropic effect is significant (yielding meaningful off-diagonal conductivity terms), we want to choose our frequency such that the losses are not too significant as to mask the guided wave propagation. Here, the operation frequency we choose is $4 THz$, corresponding to $X\approx-0.6j+0.135, X_o\approx-0.093-0.043j$.
\section{Dispersion}
\subsection{Unmagnetized tube}
In the unmagnetized case, the dispersion for each value of OAM, $n$, can be solved separately (corresponding to the matrix in equation \ref{masterMatrix} being block diagonal). Following equation (\ref{EigenmodeSingleN}), the dispersion equation can be written as
\begin{equation}
   det\left[\eta\rtwo{\sigma}\;\rtwo{M}_{E,n}- \rtwo{M}_{H,n}\right]=0.
   \label{dispersionSingleN}
\end{equation}
The dispersion curves for this case are shown using thick dotted blue lines in figure \ref{fig1}(c). There exists a degeneracy for positive and negative values of $n$, and both $n=1$ and $n=-1$ have the same dispersion. In our case, the unmagnetized tube conductivity is scalar, and therefore for $n=0$ (and only for this value), equation (\ref{dispersionSingleN}) can be completely separated to TE/TM, where only one of them will constitute the propagating mode according to the sign of the imaginary value of $\sigma$ (the sign of $X$). Here, we use $Im[\sigma]<0$ (an inductive surface), and therefore the $n=0$ mode will be a TM mode, with $B_0^i=0$. For $n\neq 0$, there is always some "mixing" between the TE and TM constituents, although a TM component will still dominate the modes due to the inductive response. 
\subsection{Magnetized tube}
When applying the magnetic bias, several effects come into play. First, $\rtwo{\sigma}$ is no longer uniform but depends on $\varphi$. This creates coupling between different values of OAM, so we can no longer separate the dispersion equations for distinct values of $n$. Consequently, each modal field distribution, derived from the eigenvector of a certain solution of the dispersion equation, will be composed of contributions from several $n$'s. Usually, significant contributions will only come from one or a few $n$, but at least theoretically, all of them are a part of the modal field. Due to this, the magnitude of the modal field will no longer be uniform across $\varphi$ (as expected in the unmagnetized case, for a simple $e^{jn\varphi}$ dependence). It is worth mentioning that this is not related to the nonreciprocal response of the magnetized tube but to the cross-sectional inhomogeneity of the conductivity matrix.  However, due to the nonreciprocity, this is expected to happen in a non-symmetric way - propagating modes with $\beta>0$ and $\beta<0$ are expected to possess differently distributed fields. This opens the possibility of directional excitation of waves by arbitrary sources, as will be shown later. 

The dispersion curve for a magnetized cylinder, with $B_0=1.5T$, is shown in solid magenta lines in figure \ref{fig1}(c), where we have neglected the associated losses to obtain a clear modal dispersion. Due to the coupling between different OAM values and the nonreciprocity, the degeneracy is now broken, and the $n=\pm 1$ branch splits into two separate branches. The $n=0$ branch, which was not degenerate in the first place, experiences some frequency shift. For this calculation, the infinite matrix dispersion equation was truncated at $n=\pm 7$, which is enough for the specific modes in our examples. Higher-order modes might require a different truncation.

The calculation of this dispersion assumes that the conductivity values did not change as a function of $k_0R$. Therefore, the vertical axis can be seen as a variation of the cylinder radius, for instance. A variation in the frequency would require that we also vary the conductivity matrix (both the diagonal and off-diagonal terms depend on frequency). An example of the surface wave dispersion when considering this dependence $\rtwo{\sigma}(\omega)$ is shown in green in figure \ref{fig1}(c). We see that qualitatively, the curves are similar, yet there is a difference in the obtained wavenumbers due to the varying conductivity values for different frequencies. The radius of the tube is taken as $a=6\mu m$ here. The dashed black line indicates the operation frequency for which we later calculate the fields in different scenarios. 
\begin{figure}[h]
\begin{center}
\noindent
  \includegraphics[width=\columnwidth]{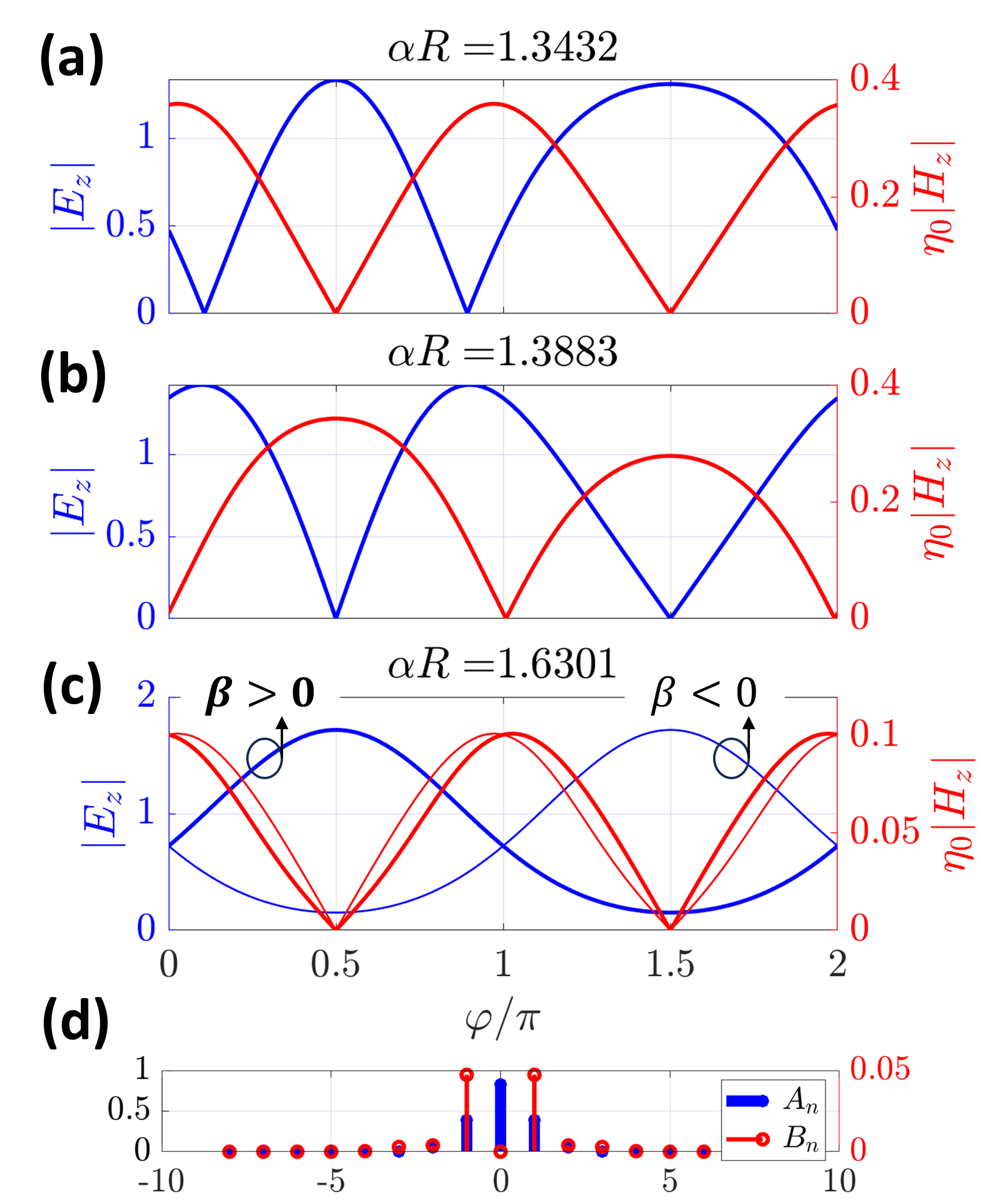}
  \caption{Mode field profiles for a magnetized graphene cylinder. (a) and (b) show the modes that split from the degenerate $n=\pm 1$ mode in the unmagnetized case. (c) The perturbed $n=0$ mode. Thin lines show the field profiles for $\beta<0$, displaying significant asymmetry. (d) The modal coefficients $A_n,B_n$ for the mode shown in (c).}
  \label{fig:GrapheneDispersionModes}
\end{center}
\end{figure}
To illustrate the modal fields, we calculate the modal coefficients. The operation point was chosen as $k_0R\approx0.5$, corresponding to a cylinder with radius $R=6[\mu m]$ in 
$f=4THz$. 
Figure \ref{fig:GrapheneDispersionModes} shows the longitudinal fields $|E_z|,|H_z|$ for the three modes that solve the presented dispersion. Panels (a) and (b) show the modes that split from the $n=\pm 1$ branch, presenting a mixed TE/TM nature. The mode close to the $n=0$ branch, shown in panel (c), is TM dominated (notice the relatively weaker $H_z$ content) since around this regime, the TE/TM coupling is weaker, a remnant of the pure TM $n=0$ mode in the unmagnetized case. The thin lines in panel (c) show the modal fields for $\beta<0$, and we see a significant difference in the field distribution compared to $\beta>0$, a clear consequence of the nonreciprocity (in the reciprocal case, we would expect a simple conjugation operation, which would preserve the profile of $|E_z|,|H_z|$). The electric field oscillates "slowly" since the $n=0$ coefficient still dominates it, but the magnetic field oscillates more rapidly since the magnetic fields are dominated by the $n=\pm 1$ coefficients, coupled through the nonuniform conductivity. This is confirmed in panel (d), where we see the coefficients composing the field corresponding to the mode shown in panel (c).

\section{Excitation of waves}
We performed full-wave simulations of different excitation schemes to illustrate the various aspects of this system using COMSOL multiphysics \cite{COMSOLref}. Throughout this section, we use the operation point previously introduced - $f=4[THz], R=6[\mu m], B_0=1.5[T]$, unless explicitly stated otherwise.
\subsection{Excitation centered around specific OAM}
\begin{figure}[h]
\begin{center}
\noindent
  \includegraphics[width=\columnwidth]{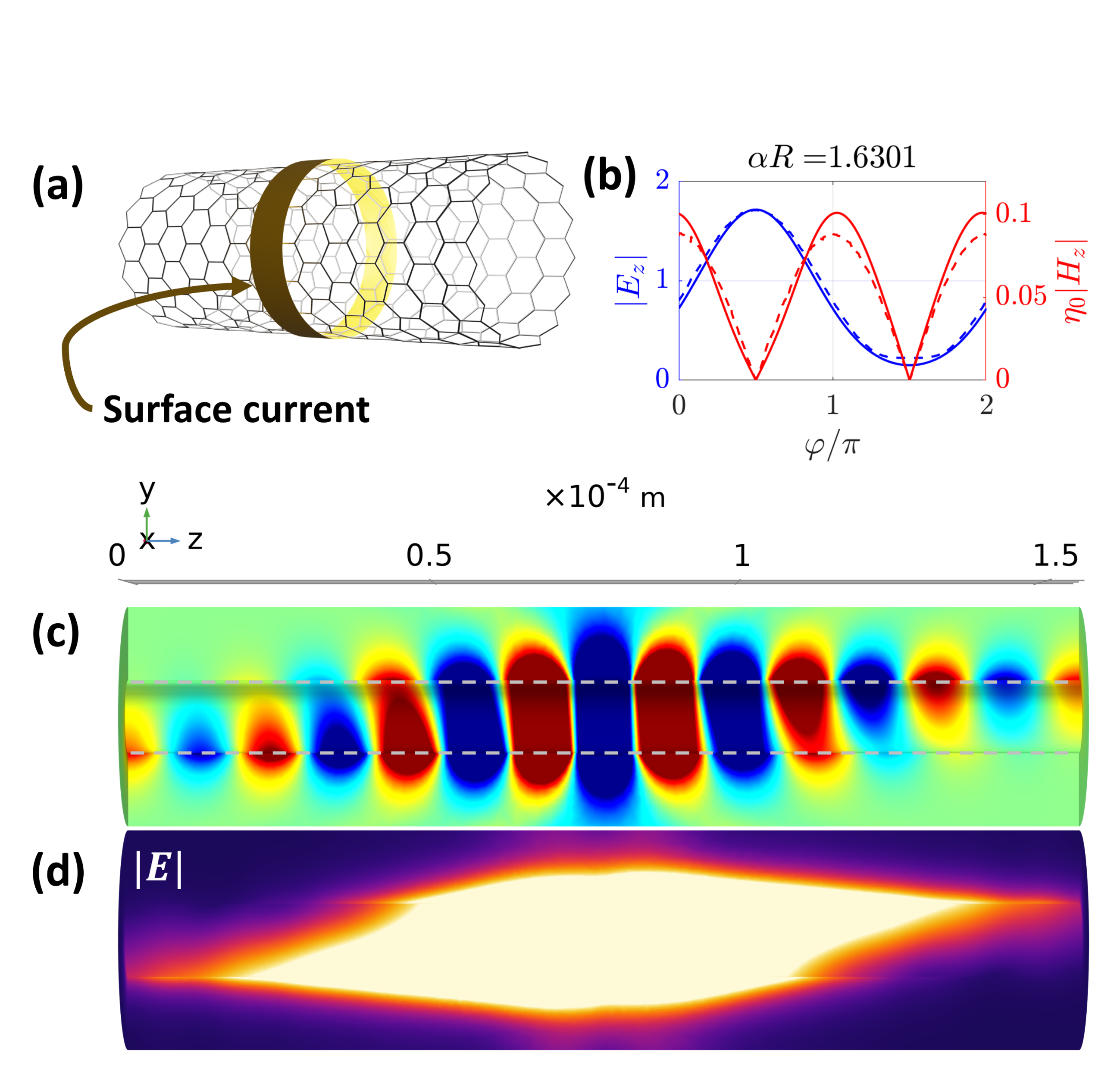}
  \caption{(a) Surface current source is placed on the yellow ring around the cylinder. (b) In dashed, the simulated $\rone{z}$ component of the propagating fields excited by the source with $n=0$, as a function of $\varphi$, measured on a circle in the XY plane with radius $1.1R$, placed $L/3$ from the source along the $+\hat{\rone{z}}$ axis. Solid lines show the analytical field distribution of the '$n=0$' mode. (c) Distribution of the excited electric field longitudinal component $E_z$. The guiding cylinder is between the dashed lines. (d) Same as (c) but for $|\boldsymbol{E}|$.}
  \label{fig:fields_n0}
\end{center}
\end{figure}

We start by inspecting the fields excited by a source centered on a specific value of $n$. To this end, we used a ring surrounding the magnetized tube, with a radius of $1.1R$, as shown in figure \ref{fig:fields_n0}(a). On this ring, we apply a surface current $\rone{J}_s=\hat{\rone{z}}e^{-jn\varphi}$ and examine the excited fields. Figure \ref{fig:fields_n0}(b) shows the profile of $E_z,H_z$, similar to Fig. \ref{fig:GrapheneDispersionModes}. The dashed line shows the distribution obtained from the full-wave simulation, and the solid line shows the analytical result for comparison. We see a good match between the field profiles since the analytically obtained electric field mainly comprises the $n=0$ component. It is worth mentioning that while the field distributions along $\varphi$ are very similar, they are not identical (in terms of the ratio between the field magnitudes, for instance). This is because we cannot excite a single mode this way, and the propagating fields contain at least some contribution from other modes.
In figure \ref{fig:fields_n0}(c), we see how propagation to the $+\hat{\rone{z}}$ direction has fields that are most dominant on the upper side (corresponding to $\varphi=\pi/2$) whereas in propagation to the $-\hat{\rone{z}}$ the lower side is more dominant. This is, again, a direct consequence of the system nonreciprocity and wouldn't occur if the system response was reciprocal (even if it is strongly dependant on $\varphi$, or even anisotropic). From figure \ref{fig:fields_n0}(c) we are also able to approximate the wavelength of the propagating wave, and it corresponds to $~\beta R\approx 1.6$ which is close to the value we extract from the analytical formulation, $\beta R\approx 1.7$. The inaccuracy is caused by the coupling of various modes in the numerical simulation. 

\subsection{Excitation by a point source}
Next, we would like to understand how the modal properties manifest when exciting the system with a point source. When using a point source, in principle, a superposition of all possible wave species is excited, each with a different amplitude. With such excitation, we can feel the system properties as a whole, essentially "probing" the Green's function. One of the consequences of nonreciprocity is that even with simple sources, we get pronounced asymmetry in the propagation of power since the field distributions of the modes that propagate in opposite directions are different. We simulated the fields excited by a dipole source $\rone{p}=p\hat{\rone{z}}$ in different locations $[x_s,y_s]$ around the center of the cylindrical waveguide, $x_s=1.1R\cos\varphi_s,y_s=1.1R\sin\varphi_s$, with $z_s$ being in the central plane, $z_s=L/2$. We then calculated the total power flowing through two cross-sections at $z=L/4$ and $z=3L/4$.
\begin{figure}[h]
\begin{center}
\noindent
  \includegraphics[width=0.9\columnwidth]{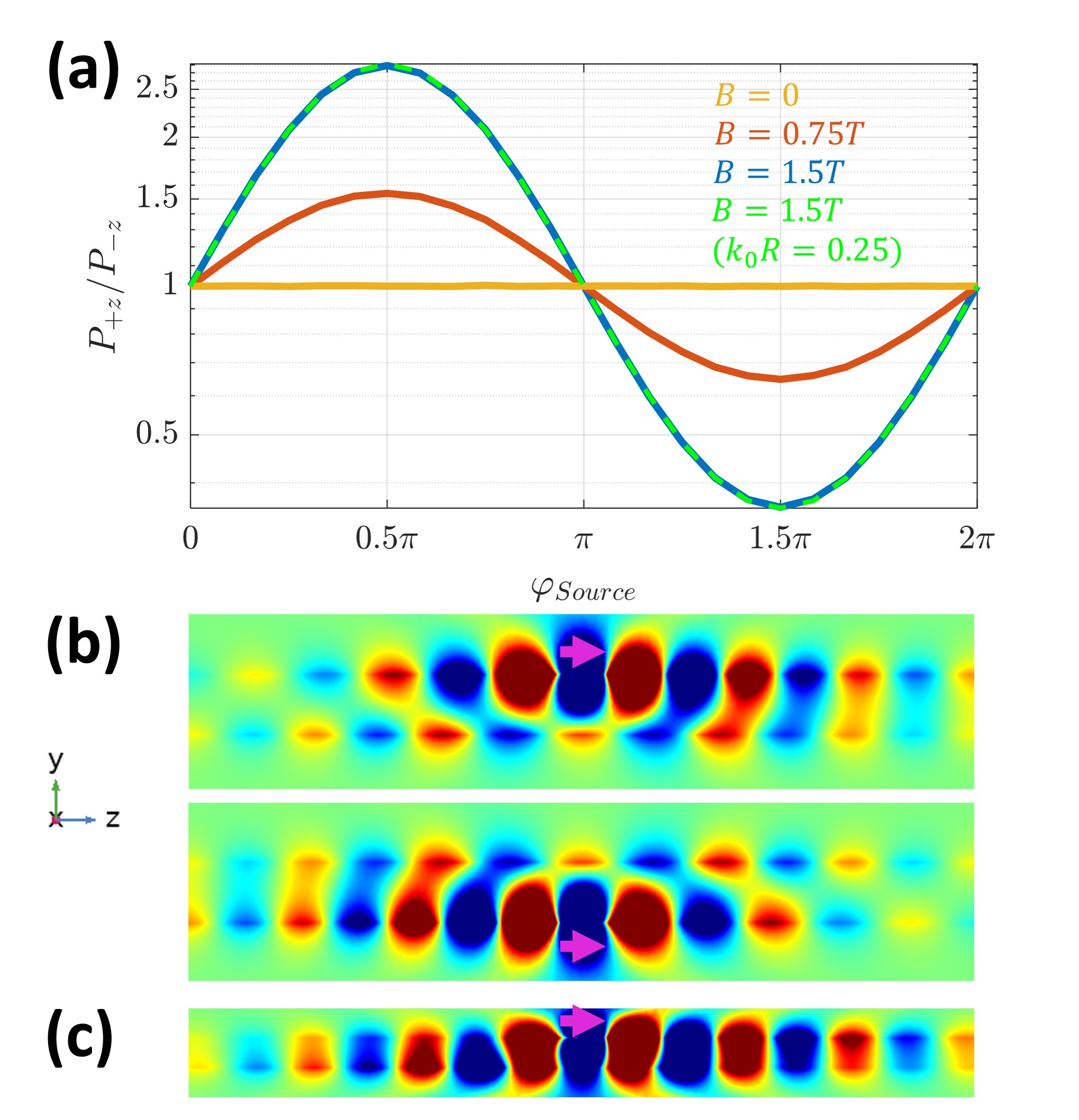}
  \caption{(a) Ratio between the total power passing through the tube cross-section to +z and -z directions, as a function of the location of the dipole source around the cylinder, $[x_s,y_s]$. (b) $E_z$ for $\phi_s=\pi/2,3\pi/2$, source dipole is shown in magenta. The asymmetry is visible. (c) $E_z$ for a smaller cylinder, with $k_0R\approx 0.25$.}
  \label{fig:fields_dip}
\end{center}
\end{figure}
Figure \ref{fig:fields_dip}(a) shows the ratio between the power that propagates to the $+z$ and $-z$ directions - the isolation ratio - as a function of the source location angle $\varphi_s$, for different values of the magnetic bias. Significant nonreciprocity is present, with $\phi_s=\pi/2,3\pi/2$ having the strongest asymmetry in power propagation. In \ref{fig:fields_dip}(b) we see the propagating fields for $\phi_s=\pi/2$ ($\phi_s=3\pi/2$) in the top (bottom) panel. The asymmetry is visible, and it is worth mentioning that this propagating power is not a result of a single propagating mode but all of the possible ones (3 in this case). Evidence of this can be seen from the "inconsistent" wavefronts in \ref{fig:fields_dip}(b), since the excited modes also have different attenuation and propagation constants. Examining the dispersion, we see that for $k_0R<~0.25$ we can obtain a "single-mode" operation regime. Figure \ref{fig:fields_dip}(c) shows the propagating fields for this case when the source is placed at $\phi_s=\pi/2$, and we can see the much more uniform wavefronts. Although the asymmetry in the field picture might not seem as pronounced, calculating the $+z$ to $-z$ propagating power reveals a similar picture, as shown in the dashed green line in figure \ref{fig:fields_dip}(a). This demonstrates that isolation is a fundamental phenomenon here and is not dependent the "mixed" excitation of modes.
\begin{figure*}[t]
\begin{center}
\noindent
  \includegraphics[width=1\textwidth]{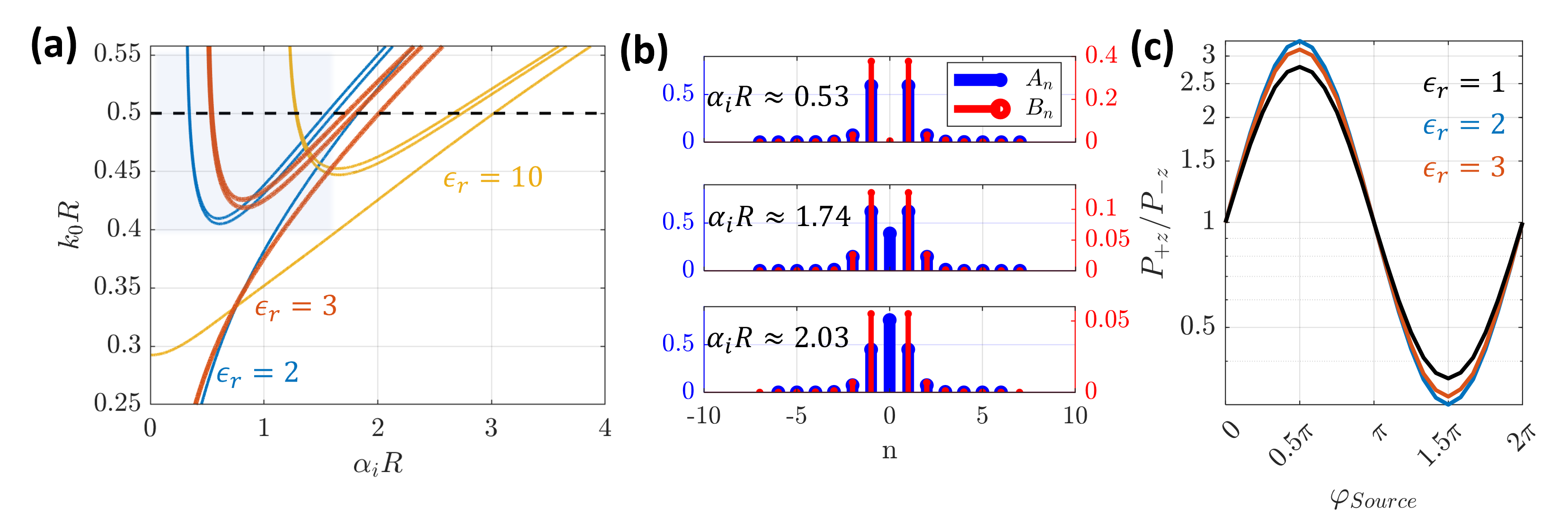}
  \caption{(a) Surface wave dispersion when the Graphene sheet is wrapped around a dielectric core for $\epsilon_{r, core}=[2,3,10]$. (b) Modal TE/TM composition for three modes (out of 5), when $\epsilon_r=3$. (c) The isolation ratio for different values of $\epsilon_r$}
  \label{fig:dispersion_dielectric_core}
\end{center}
\end{figure*}
\section{A magnetized tube wrapped around a dielectric core}
When the conductive sheet is wrapped around a dielectric core, the value of $\alpha$ in equations (\ref{eq:Ezdef}) and (\ref{eq:Hzdef}) differs between the inner and outer parts, which we will term $\alpha_{i,o}$. If the core has a dielectric constant of $\epsilon_r$ we will have $\beta^2=\alpha_i^2+\epsilon_rk_0^2$ and $\beta^2=\alpha_o^2+k_0^2$. Substituting these into the same equations and performing similar manipulations to the ones shown in appendix \ref{VaccumFields}, yields the same infinite matrix relation shown in appendix \ref{App:MatrixInf}, with the reformulated building blocks 
\begin{equation}
    \rtwo{M}_{E,n}=\begin{bmatrix}
        \frac{n\betab}{\alphab_i^2}I_n & -\frac{j\kob\sqrt{\epsilon_r}}{\alphab_i}I_n' \\
    I_n & 0
    \end{bmatrix}\;\;,\;\;\rtwo{M}_{H,n}=\rtwo{M}_{H,n}^o-\rtwo{M}_{H,n}^i,
\end{equation}
where
\begin{equation}
    \rtwo{M}_{H,n}^i=
    \frac{1}{\eta_0}
    \begin{bmatrix}
    0 & -I_n\sqrt{\epsilon_r} \\
    \frac{j\kob\sqrt{\epsilon_r}}{\alphab_i}I_n' & \frac{n\betab\sqrt{\epsilon_r}}{\alphab^2}I_n 
    \end{bmatrix},
\end{equation}
and
\begin{equation}
    \rtwo{M}_{H,n}^o=
    \frac{1}{\eta_0}
    \begin{bmatrix}
    \frac{K_n}{K_n'}\frac{n\betab\alphab_o}{j\kob}I_n\Delta & -\sqrt{\epsilon_r}\frac{\alpha_o}{\alpha_i}\frac{I_n'K_n}{K_n'} \\
    \frac{j\kob}{\alphab_o}\frac{I_nK_n'}{K_n}-\frac{(n\betab)^2}{j\kob\eta_0\alphab_o}\frac{K_nI_n}{K_n'}\Delta & \frac{n\betab\sqrt{\epsilon_r}}{\alphab_o\alphab_i}\frac{I_n'K_n}{K_n'}
    \end{bmatrix}.
\end{equation}
Here $\Delta=\left(\frac{1}{\alphab_i^2}-\frac{1}{\alphab_o^2}\right)$ which is indicative of the material contrast between the core and the surroundings, 
$I_n=I_n(\alpha_iR),K_n=K_n(\alpha_oR)$ and $I_n'=dI_n(x)/dx|_{x=\alpha_iR},K_n'=dK_n(x)/dx|_{x=\alpha_oR}$. The parameters $\eta,k$ are the intrinsic impedance and wavenumber of the dielectric medium, and $k_0,\eta_0$ are those of vacuum. One can see that when substituting $\epsilon_r=1$ we get $\alpha_i=\alpha_o,\eta=\eta_0$, and we revert back to equations (\ref{MEMHdefinition}). The solutions are obtained in the same way as in the previous case, by substituting the Graphene conductivity and $\rtwo{M}_{E,n},\rtwo{M}_{H,n}$ into equation (\ref{masterMatrix}).

Figure \ref{fig:dispersion_dielectric_core}(a) shows the dispersion relation for a dielectric core with $\epsilon_r=[2,3,10]$. The first difference we notice with respect to the all-vacuum case is that new modes emerge that did not exist previously. 
These modes are much less confined to the tube located in the shaded region in the top-left corner of the dispersion plot. We see almost no "split" in their wavenumber since the split is induced by the properties of the angularly varying conductivity, and being less confined, their interaction with the Graphene shell is much weaker. As expected, increasing the dielectric constant of the core gives larger values of $\alpha$ and shorter wavelengths of the guided waves. In figure \ref{fig:dispersion_dielectric_core}(b), we see the contribution of the different angular components. On the top, we see the new, low-confinement mode. Due to its weak interaction with the Graphene layer, there is much less coupling between different values of $n$ (we see mainly the $\pm 1$ components), and the mode is much more balanced in the TE/TM sense. In the more confined modes (bottom two panels in (b)), the interaction with the Graphene is stronger, which leads to a more pronounced TM nature and stronger coupling between different $n$. Figure \ref{fig:dispersion_dielectric_core} shows the confinement as a function of the dielectric core permittivity. We see that increasing $\epsilon_r$ may result in stronger isolation, but there is a tradeoff since increasing $\epsilon_r$ increases the mode confinement, which makes the propagating waves more prone to losses induced by the Graphene layer.
\section{Conclusions}
In this work, we have analyzed the wave propagation and the nonreciprocal characteristics of surface waves guided on a perpendicularly magnetized Graphene tube. This magnetization scheme is more realistic than the previously studied, but it results in a system with a nonuniform cross-section. The resulting propagating modes form a mixture of TE/TM components and several dominant values of OAM. The geometrical asymmetry results in more pronounced nonreciprocal effects, such as a different cross-sectional field distribution for oppositely propagating modes. These were leveraged for a directional excitation of waves using any source as a function of the source placement around the tube. Generalizing the analysis for a Graphene sheet wrapped on a dielectric core, we see new modes emerge, poorly confined and weakly interacting with the Graphene sheet, in addition to the highly confined surface waves. When exciting the waves with a point source, the dielectric core can yield more pronounced isolation, with demonstrated ratios of around $\approx 3$. This can be further improved by adding additional elements, such as an inhomogeneous core or structuring of the graphene later, which is left for future work. The continuous dependence of the isolation ratio on the direction of the magnetization lets us tailor the power propagation properties using the magnetization direction as an easy control knob without needing to solve for a different magnetization intensity or redesign the geometry and materials. This component can have several applications, such as a building block for miniature isolators and circulators in the THz range, control of signal flow, and enhance light-matter interaction in graphene-wrapped microfiber systems \cite{Silica2DReview2021}, and as a nonreciprocal antenna \cite{alu2015electrically}.
\appendix
\section{Modal field derivation}\label{VaccumFields}
Given the $\hat{z}$ components of the electric and magnetic fields in the TM and TE components respectively, we can use the following equations to derive the rest of the field components in cylindrical coordinates \cite{collin1992foundations}
\begin{equation}
    E_r=\frac{j}{\alpha^2}\left(\beta\frac{\partial E_z}{\partial r}+\frac{\omega\mu}{r}\frac{\partial H_z}{\partial \varphi}\right)
    \label{eq:Erdef}
\end{equation}
\begin{equation}
    E_{\varphi}=\frac{j}{\alpha^2}\left(\frac{\beta}{r}\frac{\partial E_z}{\partial\varphi}-\omega\mu\frac{\partial H_z}{\partial r}\right)
    \label{eq:Ephidef}
\end{equation}
\begin{equation}
    H_r=-\frac{j}{\alpha^2}\left(\frac{\omega\epsilon}{r}\frac{\partial E_z}{\partial \varphi}-\beta\frac{\partial H_z}{\partial r}\right)
    \label{eq:Hrdef}
\end{equation}
\begin{equation}
    H_{\varphi}=\frac{j}{\alpha^2}\left(\omega\epsilon\frac{\partial E_z}{\partial r}+\frac{\beta}{r}\frac{\partial H_z}{\partial \varphi}\right)
    \label{eq:Hphidef}
\end{equation}
Substituting equations (\ref{eq:Ezdef}) and (\ref{eq:Hzdef}) into these relations and applying the continuity boundary condition in equation (\ref{eq:EContinuity}) yields
\begin{equation}
    I_nA_n^i=K_nA_n^o.
\end{equation}
The continuity of $E_{\varphi}$ gives
\begin{equation}
\frac{n\beta}{\alpha^2R}\left[ I_nA_n^i-K_nA_n^o\right]-\frac{jk_0}{\alpha}\left[I_n'B_n^i-K_n'B_n^o\right]=0
\end{equation}
from which we get
\begin{equation}
    I_n'B_n^i=K_n'B_n^o
\end{equation}
In Eq. (\ref{Ddefinition}), we have defined a truncated coefficient column vector $\rone{D}_n$. Using these relations, the representation of the guided wave fields can be written as

\begin{subequations}
\begin{equation}
    E_z^i(r=R)=E_z^o(r=R)=\left[I_n,0\right]\rone{D}_ne^{-jn\varphi}e^{-j\beta z}
\end{equation}
\begin{equation}
    E_{\varphi}^i(r=R)=E_{\varphi}^o(r=R)=\left[\frac{n\beta}{\alpha^2R},-\frac{jk_0}{\alpha}I_n'\right]\rone{D}_ne^{-jn\varphi}e^{-j\beta z}
\end{equation}
\begin{equation}
    H_z^i(r=R)=\left[0,\frac{1}{\eta_0}I_n\right]\rone{D}_ne^{-jn\varphi}e^{-j\beta z}
\end{equation}
\begin{equation}
    H_z^o(r=R)=\left[0,\frac{1}{\eta_0}\frac{I_n'}{K_n'}K_n\right]\rone{D}_ne^{-jn\varphi}e^{-j\beta z}
\end{equation}
\begin{equation}
    H_{\varphi}^i(r=R)=\left[\frac{jk_0}{\alpha \eta_0}I_n',\frac{\beta n}{\alpha^2R\eta_0}I_n\right]\rone{D}_ne^{-jn\varphi}e^{-j\beta z}
\end{equation}
\begin{equation}
    H_{\varphi}^o(r=R)=\left[\frac{jk_0}{\alpha\eta_0}\frac{I_n}{K_n}K_n',\frac{\beta n}{\alpha^2R\eta_0}\frac{I_n'}{K_n'}K_n\right]\rone{D}_ne^{-jn\varphi}e^{-j\beta z}
\end{equation}
\label{eq:SurfaceFields}
\end{subequations}
These relations can be transformed into a matrix representation which will be instrumental to the convenient analysis of the guided modes, and using the wronskian identity
\begin{equation}
    I_n'(x)K_n(x)-K_n'(x)I_n(x)=\frac{1}{x}
\end{equation}
one can arrive at the definitions given in equation (\ref{MEMHdefinition}).
\section{Infinite matrix eigenmode equation}\label{App:MatrixInf}
Equation \ref{EigenmodeEq} can be written in an infinite matrix form
\begin{widetext}
\begin{equation}
    %
%
%
%
%
%
    \begin{pmatrix}
    \ddots & \vdots & \vdots & \vdots & \ddots \\
    

\hdots & \rtwo{Y}_{0}\rtwo{M}_{E,-1}-\rtwo{M}_{H,-1} & \rtwo{Y}_{-1}\rtwo{M}_{E,0} & \rtwo{Y}_{-2}\rtwo{M}_{E,1} & \hdots \\

\hdots & \rtwo{Y}_{1}\rtwo{M}_{E,-1} & \rtwo{Y}_{0}\rtwo{M}_{E,0}-\rtwo{M}_{H,0} & \rtwo{Y}_{-1}\rtwo{M}_{E,1} &  \hdots \\

\hdots & \rtwo{Y}_{2}\rtwo{M}_{E,-1} & \rtwo{Y}_{1}\rtwo{M}_{E,0} & \rtwo{Y}_{0}\rtwo{M}_{E,1}-\rtwo{M}_{H,1} &  \hdots \\


\ddots & \vdots & \vdots & \vdots & \ddots
    \end{pmatrix}
    \begin{pmatrix}
    \vdots \\
    \rone{D}_{-2} \\
    \rone{D}_{-1} \\
    \rone{D}_0 \\
    \rone{D}_1 \\
    \rone{D}_2 \\
    \vdots
    \end{pmatrix}
    = 0.
\label{masterMatrix}
\end{equation}
\end{widetext}
When a conductivity of the form given in equation (\ref{SigmaPhi}) is used, this matrix reduces to a tridiagonal block matrix.

\bibliographystyle{ieeetr}
\bibliography{Tubes}
\end{document}